\documentclass[fleqn,usenatbib]{mnras}

\usepackage{newtxtext,newtxmath}
\usepackage[T1]{fontenc}

\DeclareRobustCommand{\VAN}[3]{#2}
\let\VANthebibliography\thebibliography
\def\thebibliography{\DeclareRobustCommand{\VAN}[3]{##3}\VANthebibliography}

\usepackage{graphicx}	% Including figure files
\usepackage{amsmath}	% Advanced maths commands
\usepackage{threeparttable}
\usepackage{ae,aecompl}
\usepackage{color,soul}
\usepackage{lscape,rotating}
\usepackage{hyperref}
\usepackage[svgnames]{xcolor}

\usepackage{booktabs} 

\def\asec{\ifmmode ^{\prime\prime}\else$^{\prime\prime}$\fi}
\def\msun{M$_{\odot}$}
\def\rsun{R$_{\odot}$}

\def\degs{\ifmmode ^{\circ}\else$^{\circ}$\fi}
\def\amin{\ifmmode ^{\prime}\else$^{\prime}$\fi}
\def\asec{\ifmmode ^{\prime\prime}\else$^{\prime\prime}$\fi}
\def\fss{\hbox{$.\!\!^{\rm s}$}}        % Fractions of seconds
\def\farcs{\hbox{$.\!\!^{\prime\prime}$}}  % Fractions of arcseconds
\def\h{$^{\rm h}$}
\def\m{$^{\rm m}$}

\newcommand{\src}{J2053}
\newcommand{\fgl}{4FGL~J2054.2+6904}

\newcommand{\ergs}{erg~s$^{-1}$}
\newcommand{\flux}{erg~s$^{-1}$~cm$^{-2}$}

\newcommand{\gaia}{\textit{Gaia}}

\newcommand{\sw}{\textit{Swift}}
\newcommand{\eros}{\textit{eROSITA}}
\newcommand{\fermi}{\textit{Fermi}}
\newcommand{\kirr}{erg~cm$^{-2}$~s$^{-1}$~sr$^{-1}$}

\title[Redback pulsar PSR J2053+6905]{Discovery of the redback pulsar PSR J2053+6905 and multi-epoch optical observations of the system}
\author[Karpova et al.]{
A. V. Karpova$^1$,
J. D. Taylor$^{2,3}$,  
A. V. Bobakov$^1$,
A. Kirichenko$^{4,1}$\thanks{E-mail: aida@astro.unam.mx},
D. A. Zyuzin$^1$,   \newauthor
\ E. Fonseca$^{2,3}$,
S. V. Zharikov$^4$,
D. K. Romanov$^{1,5}$,
M. Lucio-Mendoza$^{6}$,
R. Karimov$^7$, \newauthor
\ A. Cabrera-Lavers$^{8,9}$
\\
$^1$Ioffe Institute, 26 Politekhnicheskaya, St. Petersburg, 194021,  Russia \\
$^2$Department of Physics and Astronomy, West Virginia University, P.O. Box 6315, Morgantown, WV 26506, USA \\
$^3$Center for Gravitational Waves and Cosmology, West Virginia University, Chestnut Ridge Research Building, Morgantown, WV 26505, USA \\
$^4$Instituto de Astronom\'ia, Universidad Nacional Aut\'onoma de M\'exico, Apdo. Postal 877, Baja California, M\'exico, 22800 \\
$^5$Peter the Great St. Petersburg Polytechnic University,  29 Polytechnicheskaya, St. Petersburg, 195251, Russia \\
$^6$Facultad de Ciencias F\'isico Matem\'aticas, Universidad Aut\'onoma de Coahuila. Unidad Camporredondo, Edif. A, 25000, Saltillo, Coah., México \\
$^7$Ulugh Beg Astronomical Institute, Uzbekistan Academy of Sciences, Tashkent, 100052, Uzbekistan \\
$^8$Instituto de Astrof\'isica de Canarias, V\'ia L\'actea s/n, E38200, La Laguna, Tenerife, Spain\\
$^9$GRANTECAN, Cuesta de San Jos\'e s/n, E-38712, Bre\~{n}a Baja, La Palma, Spain\\ 
}

\date{Accepted XXX. Received YYY; in original form ZZZ}

\pubyear{2026}

\begin{document}
\label{firstpage}
\pagerange{\pageref{firstpage}--\pageref{lastpage}}
\maketitle

\begin{abstract}
Using the Green Bank Telescope, we discovered the 1.68-ms binary pulsar PSR J2053+6905. 
The discovery resulted from a targeted search for radio pulsations from a redback candidate, 
previously identified through optical and X-ray observations and 
spatially coincident with the $\gamma$-ray source 4FGL J2054.2+6904.
We report the first results from the pulsar timing, including the spin period, dispersion measure, 
precisely measured Keplerian binary parameters, and the pulsar mass function.
We also found potential evidence of scintillation and long radio eclipses. 
To constrain the system parameters, we performed multi-epoch time-series photometric observations using the 2.1-m telescope at  the Observatorio Astron\'omico Nacional on the Sierra San Pedro M\'artir (Mexico) and the 1.5-m telescope of the Maidanak Astronomical Observatory (Uzbekistan), complemented by time-series optical spectroscopy with the 10.4-m Gran Telescopio Canarias. 
The optical light curves revealed a pronounced asymmetry, as well as significant long-term variations.
From the spectra, we derived a radial velocity curve semi-amplitude $K = 337(4)$ km~s$^{-1}$. 
Combining it with the pulsar mass function, we constrained the ratio of pulsar and companion masses to $M_{\rm p}/M_{\rm c} = 7.05(8)$. 
By fitting the stable segments of the light curves with a symmetric direct heating model, we derived constraints on the component masses ($M_{\rm p} \approx 2.1$~\msun\ and $M_{\rm c} \approx 0.3$~\msun), the Roche lobe filling factor, and the companion's day- and night-side temperatures. 
The long-term evolution of the light curve profiles is likely driven by the dissipation or migration of a cold stellar spot.

\end{abstract}



\section{Introduction}
\label{sec:introduction}

Redback (RB) pulsars are a subclass of the so-called `spider' pulsar systems -- close binaries with orbital periods of $\lesssim 1$~d, each comprising a millisecond pulsar (MSP) and a low-mass (0.1--1~\msun) stellar secondary \citep[e.g.][]{roberts2013}.
The latter is ablated by the pulsar wind, and the evaporated material can cause eclipses of the pulsar radio emission for a substantial part of the orbit. 
RBs are distinguished from another spider subclass, the black widows (BWs), which have very light, $\lesssim$0.05~\msun, substellar companions.
While spiders in the radio pulsar state are non-accreting, the pulsars can attain masses exceeding 2~\msun\ as a result of an earlier evolutionary phase of mass transfer from their companions.
Consequently, these systems, as part of the wider recycled neutron star population \citep[e.g.,][]{Kiziltan2013}, can provide important constraints on the equation of state of matter above the nuclear saturation density \citep[e.g.][]{kumar2023,Romani2026}.

In recent years, the population of spiders has grown significantly thanks to the \fermi\ $\gamma$-ray space telescope and targeted follow-up of unassociated sources with radio, X-ray and optical telescopes \citep[e.g.][]{Clark2023, Bangale2024, Thongmeearkom2026}.
Even in the absence of detected pulsations, promising spider candidates can be identified by investigating variability in possible X-ray and/or optical counterparts to $\gamma$-ray sources \citep[e.g.][]{braglia2020}.
In the optical band, this variability is attributed to the combination of two main mechanisms: ellipsoidal modulation due to the tidal deformation of the companion 
and its heating by the pulsar. 
While the former produces double-peaked light curves, which are observed in some RB systems \citep[e.g.][]{strader2019, Turchetta2023}, strong irradiation can lead to single-peaked profiles \citep[e.g.][]{matasanchez2023, kandel2020}.
To date, in the Galactic field approximately 90 confirmed RBs and BWs and about three dozens of candidates are known\footnote{According to the catalogue of compact binary MSPs, SpiderCat \url{https://astro.phys.ntnu.no/SpiderCAT/} \citep{spidercat}.}.

The subject of this paper, the $\gamma$-ray source \fgl, was proposed as a pulsar candidate by \citet{kerby2021}, who detected its likely X-ray/optical counterpart with \sw. 
Studies of archival optical data as well as \sw\ and \eros\ X-ray observations allowed \citet{karpova2023} to conclude that it is a promising RB binary
The Zwicky Transient Facility (ZTF) data in $g$ and $r$ bands revealed a probable orbital period of the system $P_b \approx 7.5$ h (or 0.311 d).
The light curves folded with this period show a single peak, which is characteristic of an irradiation-dominated system. 
The peak-to-peak amplitude is $\approx$0.4 mag and mean magnitudes are $r\approx 20$ and $g\approx 21$.
Analysis of the source optical spectral energy distribution (SED) provided  a radius for the companion star of 0.5$\pm$0.1~\rsun\ and an effective temperature of 5500$\pm$300 K corresponding to a G2--G9-type star.
The X-ray spectrum is hard and can be well described by a power law with a photon index of 1.0$\pm$0.3.
The unabsorbed flux in the 0.5--10 keV band is $\approx 1.7\times 10^{-13}$ \flux\ \citep{karpova2023}. 

Here, we report the discovery of the radio source PSR J2053+6905 (hereafter \src) with the Green Bank Telescope (GBT). 
This new pulsar is spatially coincident with the source of interest, confirming its RB nature.
We also carried out time-series photometric observations with the 2.1-m telescope at the Observatorio Astron\'omico Nacional on the Sierra San Pedro M\'artir (OAN-SPM) in Mexico and 1.5-m telescope of the Maidanak Astronomical Observatory (MAO) in Uzbekistan.
In addition, we performed time-series optical spectroscopy with the 10.4-m Gran Telescopio Canarias (GTC) at the Roque de los Muchachos Observatory in La Palma, Spain.
Observations and data reduction are described in Sec.~\ref{sec:data} and analysed in Sec.~\ref{sec:anaysis}.
We discuss results and make conclusions in Sec.~\ref{sec:discussion}.  

%%%%%%%%%%%%%%%%%%%%%%%%%%%%%%%%%%%%%%%%%%%%%%%%%%%%%%%%
%%%%%%%%%%%%%%%%%%%%%%%%%%%%%%%%%%%%%%%%%%%%%%%%%%%%%%%%
%%%%%%%%%%%%%%%%%%%%%%%%%%%%%%%%%%%%%%%%%%%%%%%%%%%%%%%%
%%%%%%%%%%%%%%%%%%%%%%%%%%%%%%%%%%%%%%%%%%%%%%%%%%%%%%%%
\section{observations and data reduction}
\label{sec:data}

%%%%%%%%%%%%%%%%%%%%%%%%%%%%%%%%%%%%%%%%%%%%%%%%%%%%%%%%
%%%%%%%%%%%%%%%%%%%%%%%%%%%%%%%%%%%%%%%%%%%%%%%%%%%%%%%%

\subsection{Radio data}
\label{subsec:radio}

To search for radio pulsations and confirm the nature of \src\ as a binary MSP, we performed GBT observations\footnote{Project ID GBT25B-329, PI J. D. Taylor} between October 2025 and January 2026, accumulating a total integration time of 7.5~h. 
The telescope was pointed at the coordinates of the optical counterpart candidate provided by the \gaia\ catalogue \citep{gaia2016,gaia-dr3}: R.A.(J2016) = 20\h53\m58\fss99, Dec.(J2016) = $+$69\degs05\amin19\farcs7. 
The optical position constraints are accurate to the sub-arcsecond level, which is significantly better than the X-ray localisation with an uncertainty of several arcseconds. 
Given that the angular resolution of the GBT is $\sim$10 arcmin at $\sim$1\,GHz, the inclusion of other background radio sources is possible.
However, the pulsar data processing ignores constant-in-time radio flux in search of periodic signals. Further, the orbital periods independently measured using radio and optical methods
are nearly identical, ruling out the discovery of a pulsar in a different system (see Sec.~\ref{subsec:timing}).

The observations were carried out using the L-band receiver at a centre frequency of 1500~MHz with an 800 MHz band-width and the VEGAS spectrometer `backend' operating in the incoherent filterbank pulsar mode with 4096 frequency channels. The sampling time resolution was 81.92~$\mu$s. 

The observations consisted of four sessions, each lasting 1.3--2.3~hr. 
To maximise the orbital phase coverage and mitigate the risk of missing pulsations due to potential severe radio eclipses, the sessions were scheduled to occur randomly. 
The log of the observations is presented in Table~\ref{tab:session_log}.

%%%%%%%%%%%%%%%%%%%%%% Table GBT observations %%%%%%%%%%
\begin{table}
\centering
\caption{Log of GBT Observations of \fgl.  
}
\label{tab:session_log}
\begin{tabular}{lccc}
\toprule
Session & Date       & MJD   & Exposure \\
        & yyyy-mm-dd &       & \\
\hline
1       & 2025-10-29 & 60977.2 & 2.0 h \\
2       & 2025-12-12 & 61021.8 & 1.9 h \\
3       & 2025-12-23 & 61032.4 & 1.3 h \\
4       & 2026-01-09 & 61049.0 & 2.3 h \\
\bottomrule
\end{tabular}
\end{table}
%%%%%%%%%%%%%%%%%%%%%% Table GBT observations %%%%%%%%%%

%%%%%%%%%%%%%%%%%%%%%%%%%%%%%%%%%%%%%%%%%%%%%%%%%%%%%%%%
%%%%%%%%%%%%%%%%%%%%%%%%%%%%%%%%%%%%%%%%%%%%%%%%%%%%%%%%

\subsection{Photometry}
\label{subsec:phot}

\begin{figure}
    \centering
    \includegraphics[width=\linewidth]{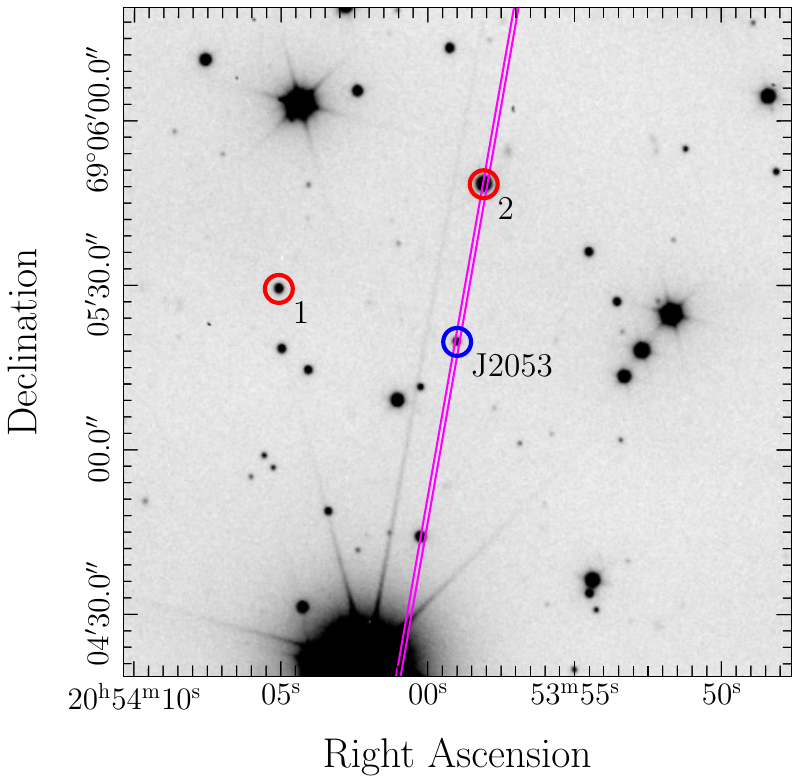}
    \caption{2$\times$2 arcmin$^2$ GTC image of the \src\ field in the $r'$ band. The position of the slit used for the spectroscopy is shown by the magenta rectangle. The \src\ optical counterpart is marked by the blue circle and the reference stars -- by red circles. }
    \label{fig:field}
\end{figure}

Time-series photometric observations of the \src\ field (Fig.~\ref{fig:field}) were carried out between 2023 and 2025 in the Johnson-Cousins $B$, $V$, and $R_c$ bands with the `Rueda Italiana' instrument mounted on the 2.1-m OAN-SPM telescope. The same set of the filters and the CCD detector were used during observational runs.
Additional observations in the Bessel $V$ filter were obtained in 2024 with the 1.5-m MAO telescope.
The weather conditions were either photometric or clear.
The observation log is presented in Table~\ref{tab:log}.
Data were reduced using the Image Reduction and Analysis Facility ({\sc iraf}) package.

The photometric calibration for OAN-SPM $V$ and $R_c$ bands was done using the SA 113-L1, SA 113-337 and SA 113-339 standards observed at the photometric night of 25 August 2025.  
The $B$-band was calibrated using field stars from the data obtained in September 2023. 
We used magnitudes of the field stars from the Panoramic Survey Telescope and Rapid Response System (Pan-STARRS) Data Release 2 catalogue \citep{flewelling2020} and converted them to 
the $B$, $V$ and $R_c$ bands using relations from \citet{2018BlgAJ..28....3K}.
We used transformation equations of the general form:
$M = m_{\rm inst} + {\rm M_{ZP}} + {\rm k_{M}} \times X + {\rm CT^{M}_{CI}} \times {\rm CI}$ where $m_{\rm inst}$  
is the instrumental magnitude of standard, ZP is the zero point, k is the extinction coefficient, $X$ is the airmass, CT is the colour term and CI is the colour index.
Using instrumental magnitudes of standards and the site extinction coefficients k$_B$ = 0.25, k$_V$ = 0.14, and k$_{R_c}$ = 0.07 \citep{schuster2, schuster1}, we calculated the zero-points 
$B_{\rm ZP} = 24.80(6)$, $V_{\rm ZP} = 24.98(1)$,   
and $R_{c,{\rm ZP}}$ = 24.87(1), and the colour terms CT$^{B}_{BV}$ = 0.07(3), CT$^{V}_{BV}$ = 0.13(2), and  CT$^{R_c}_{VR_c}$ = 0.00(2).

%%%%%%%%%%%%%%%%%%%%%% Table Photometry log %%%%%%%%%%%%
\begin{table}
\centering{
\caption{Log of the \src\ photometric observations.} 
\label{tab:log}
\begin{tabular}{cccccc}
\hline
Date        & MJD    & Filter & Exposure & Seeing,  & airmass  \\
yyyy-mm-dd  &        &        & time     & arcsec   & $\sec(z)$\\  
\hline
\multicolumn{6}{c}{2.1-m OAN-SPM} \\
2023-09-13 & 60200   & $B$    & 300 s $\times$ 3  & 1.6--1.9 & 1.5--1.8\\
           &         & $V$    & 300 s $\times$ 2  &            &           \\
           &         & $R_c$  & 300 s $\times$ 2  &            &           \\
           &         &        &                   &            &           \\
2023-09-14 & 60201   & $B$    & 300 s $\times$ 8  & 1.7--2.0   & 1.4--1.7\\
           &         & $V$    & 300 s $\times$ 7  &            &           \\
           &         & $R_c$  & 300 s $\times$ 7  &            &           \\
           &         &        &                   &            &           \\
2023-09-15 & 60202   & $B$    & 300 s $\times$ 11 & 1.4--2.0   & 1.4--1.8\\
           &         & $V$    & 300 s $\times$ 11 &            &           \\
           &         & $R_c$  & 300 s $\times$ 10 &            &           \\
           &         &        &                   &            &           \\
2023-09-16 & 60203   & $B$    & 300 s $\times$ 11 & 1.5--2.0   & 1.4--1.9\\
           &         & $V$    & 300 s $\times$ 11 &            &           \\
           &         & $R_c$  & 300 s $\times$ 11 &            &           \\
           &         &        &                   &            &           \\
2023-09-17 & 60204   & $B$    & 300 s $\times$ 14 & 1.4--1.8   & 1.3--1.8\\
           &         & $V$    & 300 s $\times$ 14 &            &           \\
           &         & $R_c$  & 300 s $\times$ 15 &            &           \\
           &         &        &                   &            &           \\
2023-09-18 & 60205   & $B$    & 300 s $\times$ 15 & 1.5--2.0   & 1.2--1.7\\  
           &         & $V$    & 300 s $\times$ 15 &            &           \\
           &         & $R_c$  & 300 s $\times$ 15 &            &           \\
           &         &        &                   &            &           \\
2024-06-08 & 60469   & $V$    & 600 s $\times$ 24 & 1.6--1.9   & 1.3--1.6\\
           &         &        &                   &            &           \\
2024-06-10 & 60471   & $V$    & 600 s $\times$ 13 & 1.3--1.9   & 1.3--1.5\\
           &         & $R_c$  & 600 s $\times$ 12 &            &           \\
           &         &        &                   &            &           \\
2024-06-11 & 60472   & $V$    & 600 s $\times$ 10 & 1.5--1.6   & 1.3--1.4\\
           &         & $R_c$  & 600 s $\times$ 10 &            &           \\
           &         &        &                   &            &           \\
2024-06-12 & 60473   & $V$    & 600 s $\times$ 8  & 1.6--1.8   & 1.3       \\
           &         & $R_c$  & 600 s $\times$ 8  &            &           \\
           &         &        &                   &            &           \\
2025-08-25 & 60912   & $V$    & 200 s $\times$ 41 & 1.4--2.0   & 1.3--1.5\\
           &         & $R_c$  & 200 s $\times$ 38 &            &           \\
\hline
\multicolumn{6}{c}{1.5-m MAO} \\
2024-06-07 & 60468   & $V$    & 360 s $\times$ 28 & 1.0--1.7  & 1.2--1.5 \\
           &         &        &                   &            &            \\
2024-06-09 & 60470   & $V$    & 360 s $\times$ 28 & 1.1--1.8   & 1.2--1.5 \\
           &         &        &                   &            &            \\
2024-06-10 & 60471   & $V$    & 360 s $\times$ 25 & 1.3--1.7   & 1.1--1.3 \\
\hline
\end{tabular}}
\end{table}
%%%%%%%%%%%%%%%%%%%%%% Table Photometry log %%%%%%%%%%%%

%%%%%%%%%%%%%%%%%%%%%%%%%%%%%%%%%%%%%%%%%%%%%%%%%%%%%%%%
%%%%%%%%%%%%%%%%%%%%%%%%%%%%%%%%%%%%%%%%%%%%%%%%%%%%%%%%

\subsection{Optical spectroscopy}
\label{subsec:spec}

Phase-resolved long-slit spectroscopy of \src\ was carried out on 2024 September 29\footnote{Program GTC8-24BMEX, PI A. Kirichenko} (MJD 60582) with the Optical System for Imaging and low-intermediate Resolution Integrated Spectroscopy\footnote{\url{https://www.gtc.iac.es/instruments/osiris+/osiris+.php}}(OSIRIS+) spectrograph installed on GTC. 
We used a 0.8 arcsec slit and the R2000B grism, which covers the 3950--5700 \AA\ range. 
The resulting spectral resolution is $\Delta \lambda = 2.3$ \AA\ in the 2$\times$2 binning mode.
We obtained 16 subsequent spectra with an individual exposure time of 1200 s covering about 75 per cent of the orbital period.

We performed standard data reduction using the {\sc pypeit} python package \citep{pypeit:zenodo,pypeit:joss_pub}. 
The wavelength calibration was carried out using a set of arc lamp spectra of Ne, Xe and HgAr, while the flux calibration was achieved using the 
G191-B2B \citep{Oke1990} spectrophotometric standard. 
However, since the slit was inclined to the parallactic angle and the airmass changes substantially over the $\approx$6.5-h duration of the observations, the absolute flux calibration introduced systematic uncertainties, such as a continuum tilt and reduced signal-to-noise ratio in the blue region ($\lesssim 4500$~\AA). 
Consequently, to eliminate these effects, all subsequent analysis was performed using normalised spectra.

%%%%%%%%%%%%%%%%%%%%% Fig. Field %%%%%%%%%%%%%%%%%%%%%%%
%%%%%%%%%%%%%%%%%%%%% Fig. Field %%%%%%%%%%%%%%%%%%%%%%%

%%%%%%%%%%%%%%%%%%%%%%%%%%%%%%%%%%%%%%%%%%%%%%%%%%%%%%%%
%%%%%%%%%%%%%%%%%%%%%%%%%%%%%%%%%%%%%%%%%%%%%%%%%%%%%%%%
%%%%%%%%%%%%%%%%%%%%%%%%%%%%%%%%%%%%%%%%%%%%%%%%%%%%%%%%
%%%%%%%%%%%%%%%%%%%%%%%%%%%%%%%%%%%%%%%%%%%%%%%%%%%%%%%%

\section{Data analysis and results}
\label{sec:anaysis}

%%%%%%%%%%%%%%%%%%%%%%%%%%%%%%%%%%%%%%%%%%%%%%%%%%%%%%%%
%%%%%%%%%%%%%%%%%%%%%%%%%%%%%%%%%%%%%%%%%%%%%%%%%%%%%%%%

\subsection{Radio searching and timing}
\label{subsec:timing}

We used the pulsar-search tools within \textsc{presto} \citep{presto2011} throughout our processing. Before searching for radio pulsations, we cleaned the data of persistent, narrow-band radio frequency interference (RFI) utilising the \textsc{rfifind} routine. Every RFI mask was manually inspected and augmented to include RFI missed by the statistical routines. 

We then constructed de-dispersed time series \citep[see, e.g.,][]{handbook2004} at trial dispersion measures (DMs) ranging from 0 to 150~pc\,cm$^{-3}$ in 3000 uniform intervals of 0.05~pc\,cm$^{-3}$. This DM range was chosen based on the distance estimate of 1.5--5.3~kpc from \citet{karpova2023}, and to account for a potentially low or high electron column density along that light of sight. The time series formed when setting $\rm{DM} = 0$ (the `topocentric times series' hereafter) was searched for apparent periodicities arising from periodic sources of RFI, local to the GBT, from the non-zero-DM periodicity search results. Only terrestrial sources have zero DM, and most MSPs are not coherently resolved before de-dispersion, so this process is not expected to remove true pulsar signals. 

The suspected close-binary nature of the system led us to conduct a `jerk search' \citep{Andersen2018_jerk} on the $\rm{DM} > 0$ time series with 8 harmonics. We chose $z_{\rm max} = 320$ and $w_{\rm max} = 140$, which are the maximum spectral drift in Fourier bins due to line-of-sight acceleration and jerk, respectively. We also divided each observation into 30~min segments to limit the observed spin-frequency drift over the course of the observation. The values of $z_{\rm max}$, $w_{\rm max}$, and the segment length were chosen to span the line-of-sight accelerations and jerks expected from binary systems with orbital periods $P_{\rm b} \approx 7.5$~h and projected semi-major axes typical of such systems in the Australia Telescope National Facility (ATNF) pulsar catalogue \citep{atnf}.\footnote{\url{https://www.atnf.csiro.au/research/pulsar/psrcat/}}

\begin{figure}
    \centering
    \includegraphics[width=\columnwidth]{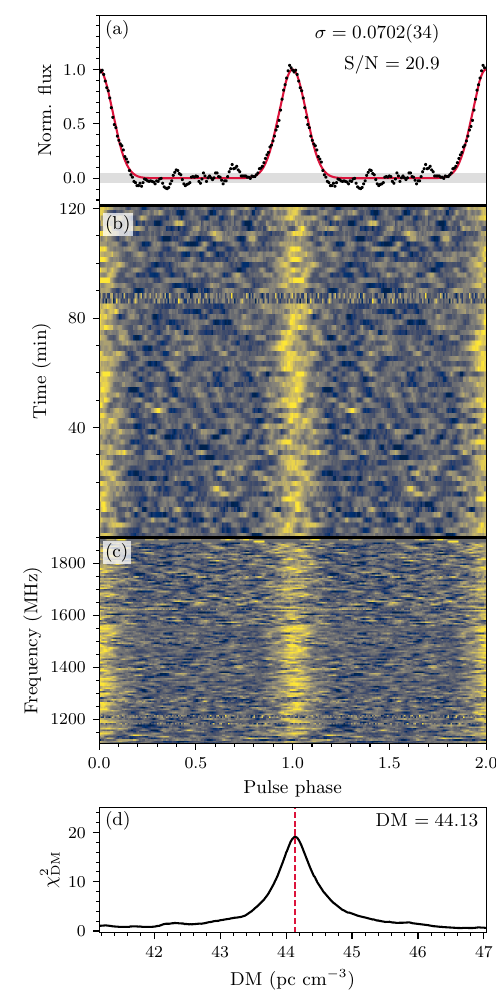}
    \caption{Detection heuristic plots from Session~1. (a) The profile normalised flux against phase folded over time and frequency. (b) Relative flux against phase over time. (c) Same as (b) but over frequency. The colour scale goes from blue to yellow for dimmest to brightest, respectively. (d) $\chi^2_{\rm DM}$ against DM when fixing the achromatic timing parameters.}
    \label{fig:fold_figure}
\end{figure}

The jerk search revealed a list of periodicity candidates ranked by the probability that they are not noise fluctuations. Candidates that displayed a $\chi^2_{\rm DM}$ with a peaked dependence on DM, typical of known radio pulsars, were folded and visually inspected, where $\chi^2_{\rm DM}$ is defined as the reduced $\chi^2$ from fitting a horizontal line to the the folded profile. PSR~J2053+6905 was the first candidate inspected with a relatively low spin period of $\sim$1.68~ms. We present its observed profile and highlighted detection heuristics from Session~1 in Fig.~\ref{fig:fold_figure}. We fit the profile against a single-component Gaussian and find a standard deviation of 0.0702(34) in units of spin phase. The pulse S/N is 20.9 when folded over the entire observation and bandwidth.

Most of the sessions show J2053 being bright across the observing time and bandwidth, though some sessions show plausible evidence of scintillation \citep[e.g.][]{Turner2024scint}. Session 4 shows strong evidence for 
extended radio eclipsing \citep[e.g.][]{Rosenthal2025eclipses} beginning roughly 20 per cent before superior conjunction, when the pulsar is furthest from the Earth along its orbit. 

In order to constrain the orbital parameters, we derived pulse times of arrival (TOAs) across all four sessions. There are roughly 40 TOAs per session for Sessions 1, 2, and 3, and 5 TOAs in Session 4. We lack more TOAs in Session 4 because the pulsar becomes too dim to be detected at all starting at around 2~ks into the observation, likely due to a radio eclipse. We assumed zero eccentricity, and fit for the spin period $P_s$, binary orbital period $P_b$, the projected pulsar's semi-major axis $x=a_{\rm p}\sin{(i)} /c$, and the pulsar's epoch of ascending node $T_{\rm asc}$. $a_{\rm p}$ is the true semi-major axis of the pulsar with respect to the system centre of mass, $i$ is the orbital inclination, and $c$ is the speed of light in a vacuum.
We attempted to phase connect the observations together using the software \textsc{aptb} \citep{Taylor2024APTB}, which was designed to phase connect binary millisecond pulsars with disparate data. The algorithm found hundreds of separate ``solutions" with nearly identical $\chi^2$, where a solution is the identification of the number of neutron star rotations between observation sessions. Therefore, we conclude that we cannot uniquely phase connect this pulsar with the data available. To remedy this, we are planning observations with the CHIME\footnote{Canadian Hydrogen Intensity Mapping Experiment} radio telescope \citep{CHIME2021} and the GBT to be presented in a forthcoming work. Thus, for now, we must also fit for arbitrary phase offsets between observations, known as `jumps' \citep[see e.g.][]{Taylor2024APTB} in order to most strongly constrain the pulsar's timing parameters. 

\begin{figure}
    \centering
    \includegraphics[width=\columnwidth]{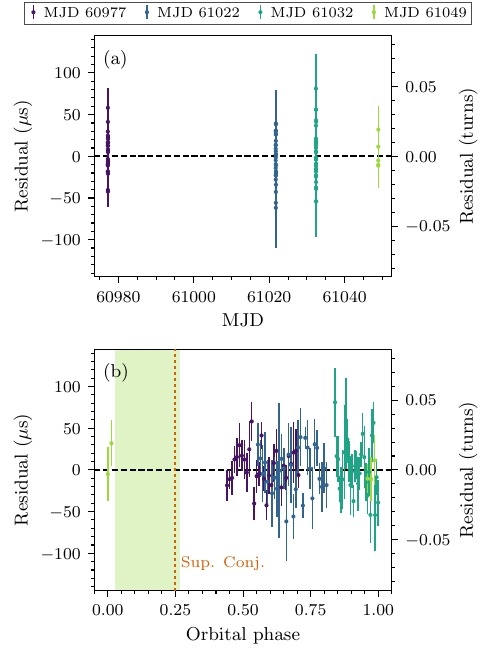}
    \caption{Timing residuals for J2053 over time (a) and orbital phase (b). We define the orbital phase as the mean anomaly referenced to $T_{\rm asc}$ divided by $2\pi$. The right-hand axes give the same residuals in rotational phase. Colours mark the four observing epochs in chronological order (MJD 60977, 61022, 61032, 61049; dark to light). The dotted vertical line of panel (b) marks superior conjunction. The light green vertical bar marks the orbital phase where J2053 was too dim to be observed during GBT Session~4.}
    \label{fig:residuals}
\end{figure}

The results of our fit are presented in Table~\ref{tab:psr_pars}.
Our results allow us to calculate the pulsar mass function,
\begin{equation}
    f_{\rm p} = \frac{M_{\rm c}^3\ \text{sin}^3i}{(M_{\rm p} + M_{\rm c})^2} = \frac{x^3}{T_\odot}\frac{2\pi}{P_b} = 0.00352186(22)~{\rm M_\odot},
\end{equation}
where $M_{\rm p}$ and $M_{\rm c}$ are the masses of the pulsar and companion, respectively, and $T_\odot \approx 4.925490947641~\rm{\mu s}$ is the nominal solar mass parameter.
We show the residuals of the fit in Fig.~\ref{fig:residuals} with respect to time and orbital phase. We scaled the TOA uncertainties by a constant factor (known as `EFAC') so that the reduced $\chi^2$ became unity. This scales the uncertainties on the timing parameters to account for an unknown noise source. Before scaling, the reduced $\chi^2$ was $\sim$1.29. The light green vertical bar marks the extent of Session 4 that has GBT data but no detected pulsations. The remaining orbital phase without TOAs is due to a true lack of observational coverage across every session.

%%%%%%%%%%%%%%%%%%%%% Table Rado parameters %%%%%%%%%%%%
\begin{table}
\renewcommand{\arraystretch}{1.2}
\caption{Parameters of \src\ from radio observations.}
\label{tab:psr_pars}
\begin{center}
\begin{threeparttable}
\begin{tabular}{ll}
\hline
Intrinsic spin period $P_s$, ms                         & 1.6755356257(69) \\
Projected semi-major axis $x=a_{\rm p}\sin i/c$, lt-s & 0.682187(14) \\
Orbital period $P_b$, h                               & 7.46657907(35) \\
Time of ascending node $T_{\text{asc}}$, MJD TDB      & 60977.0508876(27) \\
Reduced $\chi^2$ before scaling errors & $\sim$1.29\\
Post-fit weighted RMS, $\mu$s & $\sim$24\\
Mass function $f_{\rm p}$, \msun                      & 0.00352186(22) \\
Minimal companion mass $M_{\rm c,min}$, \msun         & 0.2 \\
Dispersion measure DM, pc cm$^{-3}$                   & 44.13 \\
DM distance $D_{\rm NE2001}$, kpc                     & 2.3 \\
DM distance $D_{\rm NE2025}$, kpc                     & 3.0 \\
DM distance $D_{\rm YMW16}$, kpc                      & 2.7 \\
\hline
\end{tabular}
\footnotesize{\textit{Notes.}
Numbers in parentheses denote 1$\sigma$ uncertainties related to the last significant digits. The uncertainties were calculated after the TOA uncertainty scaling.
$M_{\rm c,min}$ is calculated assuming the pulsar mass of 1.35~\msun\ and inclination of 90\degs.
$D_{\rm NE2001}$, $D_{\rm NE2025}$ and $D_{\rm YMW16}$ are DM distances derived using the NE2001 \citep{ne2001}, NE2025 \citep{ne2025} and YMW16 \citep*{ymw2016} models for the distribution of free electrons in the Galaxy, respectively. 
}
\end{threeparttable}
\end{center}
\end{table}
%%%%%%%%%%%%%%%%%%%%% Table Rado parameters %%%%%%%%%%%%

The derived orbital period is in excellent agreement with the value obtained from archival optical data (\(P_b = 7.4666(9)\)~h; \citealt{karpova2023}), firmly establishing the association of the X-ray/optical source with the pulsar.
Henceforth, we adopt the $P_b$ and $T_{\rm asc}$ as measured by radio timing to define the orbital phases for our photometric and spectroscopic analyses (under this convention, phase 0.25 corresponds to the superior conjunction of the pulsar and thus the inferior conjunction of the companion star). 
A more comprehensive timing analysis, including both these radio data and potential gamma-ray pulsation detections, will be presented in a forthcoming paper.

%%%%%%%%%%%%%%%%%%%%%%%%%%%%%%%%%%%%%%%%%%%%%%%%%%%%%%%%
%%%%%%%%%%%%%%%%%%%%%%%%%%%%%%%%%%%%%%%%%%%%%%%%%%%%%%%%

\subsection{Optical light curves}
\label{subsec:lc}

%%%%%%%%%%%%%%%%%%%%% Fig. LCs %%%%%%%%%%%%%%%%%%%%%%%%%
\begin{figure*}
\centering
\includegraphics[width=0.7\linewidth]{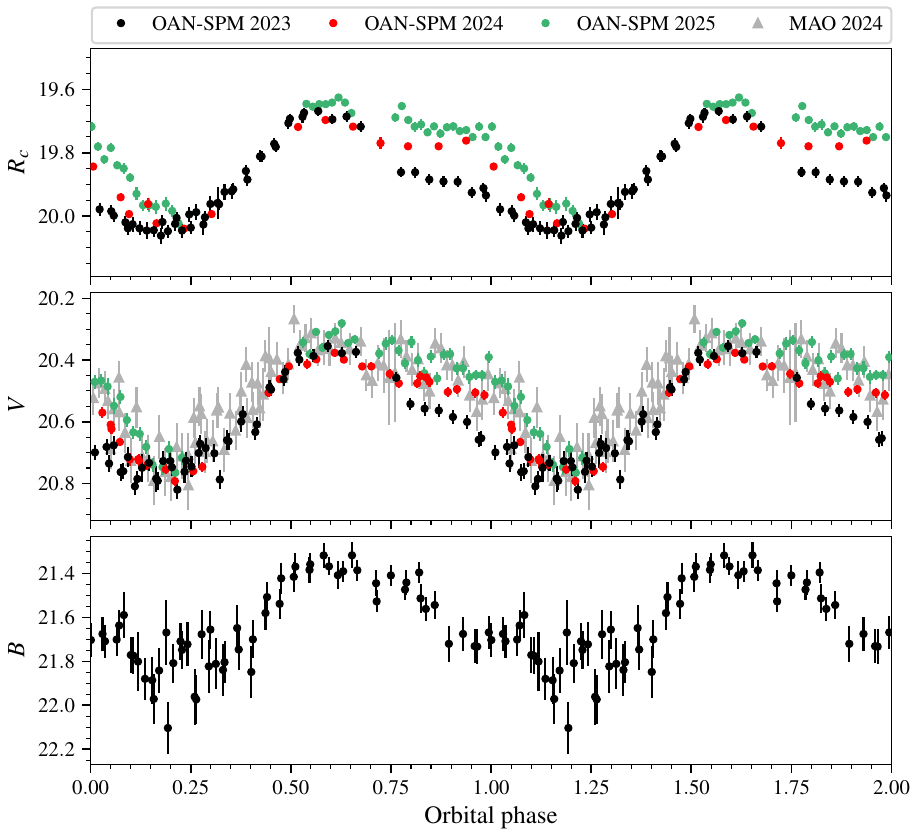}
\caption{$B$-, $V$- and $R$-band light curves of the \src\ companion folded with the orbital period (two orbital cycles are shown for clarity). 
Data from different epochs and telescopes are marked by various colours and symbols.}
\label{fig:lc}
\end{figure*}
%%%%%%%%%%%%%%%%%%%%% Fig. LCs %%%%%%%%%%%%%%%%%%%%%%%%%

The obtained optical light curves folded with the orbital period are presented in Fig.~\ref{fig:lc}. 
They demonstrate a strong asymmetry relative to the orbital phase $\phi = 0.75$ which was not evident in the noisier ZTF data (see fig.~3 of \citealt{karpova2023}). 
Such a shape cannot be described by simple direct heating models or ellipsoidal modulation, which are often used to model spider system light curves. 
Moreover, the observations revealed significant variations in the light curve shape from epoch to epoch.
Interestingly, these variations are highly phase-dependent: while the segment spanning phases $\phi \sim 0.25$--$0.75$ remains stable across all observed epochs, the remaining part of the light curve shows a substantial
brightness increase from 2023 through 2025.

%%%%%%%%%%%%%%%%%%%%%%%%%%%%%%%%%%%%%%%%%%%%%%%%%%%%%%%%
%%%%%%%%%%%%%%%%%%%%%%%%%%%%%%%%%%%%%%%%%%%%%%%%%%%%%%%%

\subsection{Spectra and radial velocities}
\label{subsec:rad-vel}

%%%%%%%%%%%%%%%%%%%%% Fig. 1-D spectra %%%%%%%%%%%%%%%%%
\begin{figure*}
\centering
\includegraphics[width=1.0\linewidth]{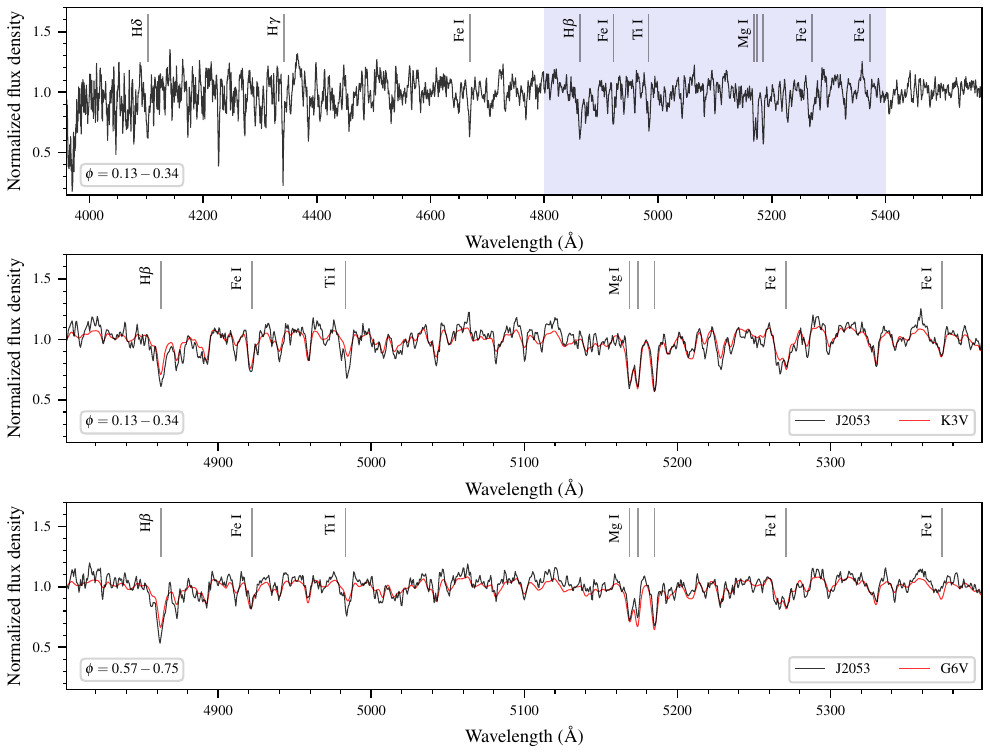}
\caption{
\textit{Top}: Normalised spectrum of the \src\ companion, obtained by co-adding four spectra taken near the orbital phase of minimum brightness. 
The shaded area indicates the spectral range zoomed-in in the lower panels. 
It contains the most prominent absorption lines, including the \ion{Mg}{i} triplet, which was used for the radial velocity measurements (see text).
\textit{Middle}: A close-up view of the shaded region (black line), overlaid with a K3V stellar template (red line) from \citet{Kesseli2017} for comparison. 
\textit{Bottom}: The same spectral region near the maximum brightness (black line) compared with a G6V template (red line).
The phase intervals for the co-added frames and positions of relatively strong absorption lines are indicated in the panels.
}
\label{fig:spec}
\end{figure*}
%%%%%%%%%%%%%%%%%%%%% Fig. 1-D spectra %%%%%%%%%%%%%%%%%

%%%%%%%%%%%%%%%%%%%%% Fig. Trailed spectrum %%%%%%%%%%%%

\begin{figure*}
\centering
\includegraphics[width=1.0\linewidth]{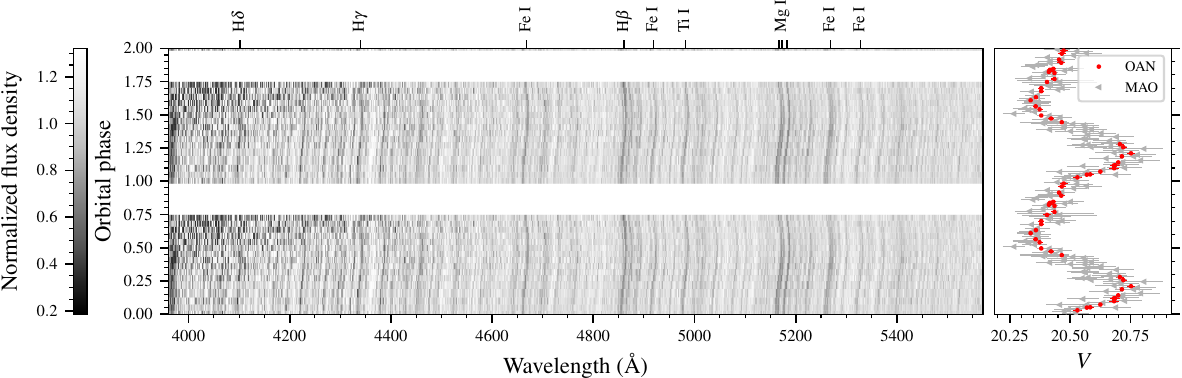}
\caption{
\textit{Left}: Normalised trailed spectrum of \src. 
Positions of the relatively strong absorption lines are marked.
Two orbital cycles are shown for clarity.
\textit{Right}: $V$-band light curves obtained with OAN-SPM and MAO in 2024 (the closest epoch to the spectral observations).
}
\label{fig:trailed}
\end{figure*}
%%%%%%%%%%%%%%%%%%%%% Fig. Trailed spectrum %%%%%%%%%%%%

%%%%%%%%%%%%%%%%%%%%% Fig. RVs %%%%%%%%%%%%%%%%%%%%%%%%%
\begin{figure}
\centering
\includegraphics[width=1\linewidth]{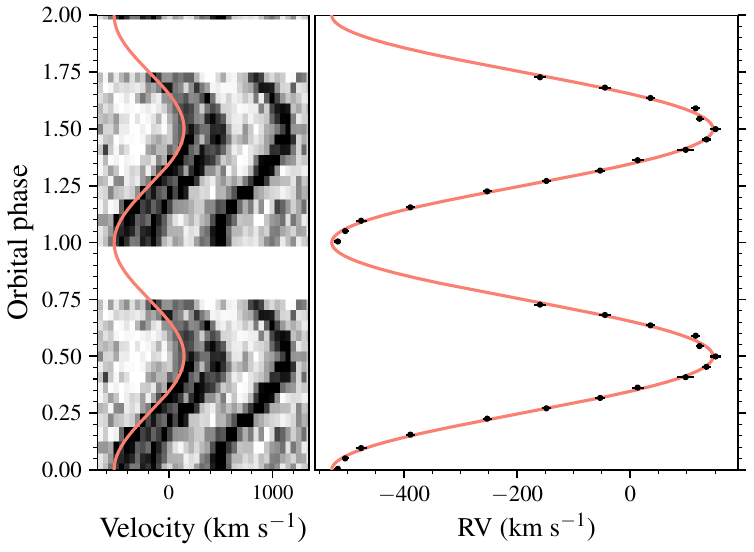}
\caption{
Trailed spectrum of the \ion{Mg}{i} triplet (left) and the RV curve of the \src\ companion (black dots with error bars; right).
The orange line in both panels is the best-fitting model.
Two orbital cycles are shown for clarity.
}
\label{fig:rv}
\end{figure}
%%%%%%%%%%%%%%%%%%%%% Fig. RVs %%%%%%%%%%%%%%%%%%%%%%%%%

The top panel of Fig.~\ref{fig:spec} demonstrates the normalised spectrum of the \src\ counterpart candidate obtained near the minimum brightness phase. 
For comparison, the middle and bottom panels show zoomed-in sections of this spectrum and the spectrum near the maximum brightness phases. 

In Fig.~\ref{fig:trailed}, we present the trailed spectrum of \src. 
The positions of the absorption lines clearly show sine-like variations with the orbital phase, indicating a significant radial velocity (RV) amplitude.
To measure this amplitude, we applied the following procedure.
We used the \ion{Mg}{i} triplet lines (5168.76, 5174.13, and 5185.05~\AA\ in vacuum), which are the brightest features in the trailed spectrum and are clearly distinguished at all phases\footnote{We also note that metallic lines, in general, and \ion{Mg}{i} lines, in particular, provide the most reliable RV measurements for irradiated companions \citep{Linares2018}.}. 
For each observed spectrum, we normalised the continuum around the triplet using a linear fit to line-free regions.
The normalised profiles were subsequently modelled using a combination of three Gaussian components.
The free parameters were the central wavelength of the first component (which determines the RV), the common line width, and the depths of the three lines.

To estimate uncertainties, we generated 2000 synthetic spectra by adding Gaussian noise to the original absolute fluxes based on their corresponding error arrays. 
For each simulated spectrum, the local continuum was independently re-normalised before re-fitting the profiles.
We refitted each simulated spectrum and computed the mean and the standard deviation of the resulting distribution to obtain the RVs and their statistical uncertainties. 

Crucially, to check for potential systematic effects, we analysed a brighter reference star (star~2 in Fig.~\ref{fig:field}). 
This is a G-type star according to \gaia\ catalogue, and the \ion{Mg}{i} triplet is clearly detected in its spectra. 
Consequently, we applied the same RV measurement technique as for \src\ and discovered a significant temporal drift with a peak-to-peak amplitude of $\Delta\text{RV} \approx 30$~km~s$^{-1}$. 
The drift follows a smooth, non-monotonic trend and cannot not be attributed to the possible binary motion\footnote{To verify the drift, we checked one more star (Gaia DR3 2271095628575544448), located on the opposite side of the slit from the target and found that it exhibited the same trend despite the lower signal-to-noise ratio. 
It can be caused by a combination of residual instrumental flexure, differential atmospheric refraction, and tracking or slit-positioning variations (see, e.g., \citealt{Moiseev2008}).}.
Because the true RV of the reference star was unknown, we used its mean RV across all frames as the stationary baseline. 
The deviations from this mean in each frame reflect this drift. 
We then subtracted these deviations directly from the measured RVs of \src\ to correct for the systematic shift. 
Finally, the statistical uncertainties of the reference star's RV measurements were added in quadrature to those of \src.

The resulting RV curve was then fitted with a function:
\begin{equation}
\text{RV}(t) = \gamma - K \text{cos}\left(2\pi\frac{t - T_{\rm asc}}{P_b}\right),    
\end{equation}
where $K$ is the RV curve semi-amplitude and $\gamma$ is the systemic velocity.
An initial fit using only the statistical errors yielded $\chi^2/\text{d.o.f.} = 23/14$. 
We thus added a systematic uncertainty component in quadrature to the statistical errors of each data point to achieve a reduced $\chi^2_\nu = 1$.
Using this conservative approach, we derived $K = 337\pm 4$ km~s$^{-1}$ and $\gamma = -191\pm 3\ ({\rm stat.}) \pm 9\ ({\rm sys.})$ km~s$^{-1}$.
The systematic uncertainty of $\gamma$ is the standard deviation of the reference star's measured RVs.
The RV curve and its best fit are shown in Fig.~\ref{fig:rv}.

Now we can estimate the companion mass function:
\begin{equation}
f_{\rm c} = \frac{M_{\rm p}^3\ \text{sin}^3i}{(M_{\rm p} + M_{\rm c})^2} = \frac{P_b K^3}{2 \pi G} = 1.235(40)~{\rm M_\odot},    
\end{equation}
where $G$ is the gravitational constant.
Combining this with the mass function from the radio timing (Table~\ref{tab:psr_pars}), we obtained the mass ratio 
$M_{\rm p}/M_{\rm c} = 7.05(8)$.

%%%%%%%%%%%%%%%%%%%%%%%%%%%%%%%%%%%%%%%%%%%%%%%%%%%%%%%%
%%%%%%%%%%%%%%%%%%%%%%%%%%%%%%%%%%%%%%%%%%%%%%%%%%%%%%%%

\subsection{The system parameters}
\label{subsec:lc-fit}

%%%%%%%%%%%%%%%%%%%%% Fig. RVs %%%%%%%%%%%%%%%%%%%%%%%%%
\begin{figure}
\centering
\includegraphics[width=1\linewidth]{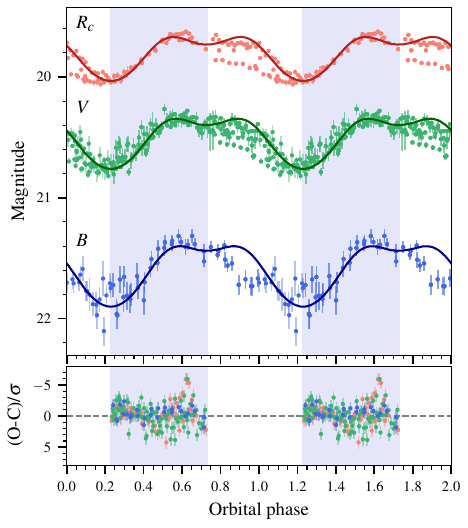}
\caption{\textit{Top}: $B$-, $V$- and $R$-band light curves of the \src\ companion folded with the orbital period and the best-fitting model shown by the solid lines (two orbital cycles are presented for clarity). 
The shaded areas mark the stable sections of the light curves selected for fitting (see text). 
\textit{Bottom}: Residuals calculated as the difference between the observed (O) and calculated (C) magnitudes for each data point in terms of the magnitude error ($\sigma$).
}
\label{fig:lc-fit}
\end{figure}
%%%%%%%%%%%%%%%%%%%%% Fig. RVs %%%%%%%%%%%%%%%%%%%%%%%%%

%%%%%%%%%%%%%%%%%%%% Table LC fit results %%%%%%%%%%%%%%%%%%
\begin{table}
\renewcommand{\arraystretch}{1.2}
\caption{The light curve fitting results for \src.}
\label{tab:fit} 
\begin{center}
\begin{threeparttable}
\begin{tabular}{lc}
\hline
\multicolumn{2}{c}{Fitted parameters}                     \\
\hline
Pulsar mass $M_{\rm p}$, \msun                     & 2.12(11) \\
Distance $D$, kpc                                  & 3.6(1)\\
Reddening $E(B-V)$                                 & 0.24(2) \\
Night-side temperature $T_{\rm n}$, K              & 4960(100) \\
Inclination$^*$ $i$, deg                           & 66(4)\\
Roche lobe filling factor $f_x$                    & 0.88(5) \\
Irradiation factor $K_{\rm irr}$, $10^{20}$ \kirr  & 1.18(24)       \\
\hline
$\chi^2$/d.o.f.                                    & 469/149 \\
\hline
\multicolumn{2}{c}{Derived parameters$^\dag$ }                   \\
\hline
Companion mass $M_{\rm c}$, \msun                  & 0.30 \\
Companion radius $R_{\rm c}^x$, \rsun              & 0.72 \\
Companion radius $R_{\rm c}^y$, \rsun              & 0.52 \\
Maximal day-side temperature $T_{\rm d}$, K        & 5550 \\
\hline
\end{tabular}
\footnotesize{\textit{Notes.}
Numbers in parentheses denote the 1$\sigma$ uncertainties in the last quoted digits.
$R_{\rm c}^{x}$ and $R_{\rm c}^{y}$ are the radii of the ellipsoidal companion. 
The former is measured along the line connecting the centres of the binary components. \\
$^*$ The allowed range of orbital inclinations is determined by the uncertainties in the RV semi-amplitude and the pulsar mass.   \\
$^\dag$The reported values correspond to the best-fitting model parameters.  
}
\end{threeparttable}
\end{center}
\end{table}
%%%%%%%%%%%%%%%%%%%% Table LC fit results %%%%%%%%%%%%%%%%%%

To estimate the system parameters -- including the binary inclination, pulsar mass, and companion properties -- we fitted the folded $B$-, $V$- and $R$-band light curves using a symmetric direct heating model, which assumes blackbody emission from the companion surface. 
A detailed description of the model is given by \citet{zharikov2013, zharikov2019}. 
For the fitting procedure, we considered only the orbital phase interval from 0.23 to 0.73,  
where all data sets exhibit consistent behaviour in different epochs. 
In contrast, the remaining phase intervals were excluded due to significant epoch-to-epoch variations in the light-curve shape, which indicate time-dependent asymmetries unaccounted for by the model.
The free parameters of the fit are the distance to the binary system $D$, the interstellar reddening $E(B-V)$, the pulsar mass $M_{\rm p}$, the orbital inclination $i$, the irradiation efficiency factor $K_{\rm irr}$, which characterises the heating of the companion, the Roche-lobe filling factor of the companion $f$ and the companion's night-side effective temperature, $T_{\rm n}$.

We fixed the orbital period at the value derived from the radio timing (Table~\ref{tab:psr_pars}) and the calculated mass ratio.  
The companion's night-side temperature, the irradiation factor, the Roche-lobe filling factor, and the orbital inclination were initialised using a random sampling procedure. 
The initial pulsar mass was set to the canonical 1.4~\msun\ (used as an optimisation starting point, not a statistical prior).
The minimum of the $\chi^2$ function
\begin{equation}
\chi^2 = \sum_i^{N_{BVR}} \left(\frac{m^{\rm obs}_i-m^{\rm mod}_i}{\sigma_{m^{\rm obs}_i}}\right)^2 
\end{equation}
was determined using a gradient-descent algorithm, which was adopted because it significantly reduces the computational cost of the optimisation. 
The uncertainties of the model parameters  were estimated following the method of \citet{1976ApJ...208..177L}. 
The results of the fit are presented in Table~\ref{tab:fit}, and the corresponding best-fitting model is shown in Fig.~\ref{fig:lc-fit}. 
Although the reduced $\chi^2$ is relatively high, likely reflecting additional variability not accounted for by the model, the derived parameters are physically plausible. 
In particular, we find good agreement between the spectral type of the companion inferred from spectroscopy and its temperatures derived from the light-curve modelling. 
Indeed, a spectral classification, performed via a $\chi^2$ minimisation routine to match individual rest-frame normalised spectra against a grid of stellar templates from \citet{Kesseli2017} within the 4500--5500~\AA\ range, yielded a best-match spectral type of $\sim$K2V--K3V near the minimum and $\sim$G6V near the maximum brightness phases (see Fig.~\ref{fig:spec}).
Extending the orbital phase range included in the light-curve fitting results in higher \(\chi^2\) values due to the larger dispersion of the data, but does not significantly affect the best-fitting solution.

%%%%%%%%%%%%%%%%%%%%%%%%%%%%%%%%%%%%%%%%%%%%%%%%%%%%%%%%
%%%%%%%%%%%%%%%%%%%%%%%%%%%%%%%%%%%%%%%%%%%%%%%%%%%%%%%%
%%%%%%%%%%%%%%%%%%%%%%%%%%%%%%%%%%%%%%%%%%%%%%%%%%%%%%%%
%%%%%%%%%%%%%%%%%%%%%%%%%%%%%%%%%%%%%%%%%%%%%%%%%%%%%%%%

\section{Discussion and conclusions}
\label{sec:discussion}

The detection of radio pulsations, with an orbital period perfectly matching the optical modulation, confirms the RB nature of PSR J2053+6905.
Notably, it is among the fastest spinning MSPs currently known. 
Its spin period makes it the tenth fastest pulsar known according to the ATNF catalogue.
Furthermore, it represents the third fastest RB discovered to date, trailing only behind PSR~J1748$-$2446ad in the globular cluster Terzan~5 ($P = 1.396$~ms; \citealt{Hessels2006}), which remains the fastest spinning pulsar, and PSR~J1859+0313g ($P = 1.61$~ms; \citealt{wang2025}).

The \src\ optical light curves exhibit a pronounced, long-term asymmetry. 
Interestingly, this variability is phase-dependent: the segment spanning phases $\phi \sim 0.25$--$0.75$ remains relatively stable across observed epochs, whereas the remaining part of the light curve shows a substantial brightness increase from 2023 through 2025.
Since our photometry was performed in three distinct observational epochs separated by approximately one year, these variations indicate only long-term evolution of the light curve shape, but cannot rule out variability on shorter timescales (such as weeks or months).

Asymmetric light curves in spider binaries are often modelled via non-uniform heating patterns, such as those produced by an asymmetric intrabinary shock (IBS) geometry \citep{romani&sanchez2016} or the ducting of relativistic IBS particles along the companion's magnetic poles (IBS-B model) \citep{sanchez&romani2017}. 
Note that the pure IBS model seems unable to describe such a strong optical asymmetry for \src\ while the IBS-B model produces localised hot spots, which would introduce an excess flux over the baseline profile which is difficult to attribute to the \src\ case.
Alternative models involve heat redistribution over the companion surface via global convection and diffusion \citep{kandel&romani2020, voisin2020} which can skew and shift light curve profiles. 
However, these models are expected to be relatively stable.

The most plausible explanation for the observed photometry evolution of \src\ seems to involve a dynamic reconfiguration of the companion's photosphere.
Since the secondary is tidally locked and thus spinning exceptionally fast, one can expect an extremely powerful magnetic dynamo capable of generating long-lived stellar spots \citep[e.g.,][]{clark2021}. 
Thus, one can suggest the slow dissipation or migration of a giant, prominent cold spot(s) to explain the multi-year brightness increase. 
Such spots were previously invoked to interpret the observational properties of, e.g. PSR J2039$-$5617 \citep{clark2021} or PSR J1431$-$4715 \citep{deMartino2024}.

The stable part of the light curve is driven by a combination of ellipsoidal modulation and pulsar heating. 
To reproduce the observed light curves, only modest heating of the companion is required, resulting in a day-side temperature of about 5550~K. 
This value agrees well with the temperature of $5500 \pm 300$~K obtained independently from the SED fitting \citep{karpova2023}. 
The derived night-side temperature of $4960 \pm 100$~K is typical for RB systems, which generally exhibit temperatures in the range of 4000--6000~K \citep{Turchetta2023}. 
The resulting day-to-night temperature difference of $\approx 600$~K is also common for this class of objects. 

The \src\ proper motion  from the \gaia\ catalogue, $\mu \approx 7.7$ mas~yr$^{-1}$, corresponds to a transverse velocity of 131 km~s$^{-1}$ at the distance of 3.6 kpc (Table~\ref{tab:fit}).
Combining this with the systemic radial velocity of $\gamma = -191$ km~s$^{-1}$ yields a total velocity of 232 km~s$^{-1}$.
This is a rather high value for recycled pulsars \citep{hobbs2005, Shamohammadi2024pulsarvel}.

Taking the $\gamma$-ray \citep{4fgl-dr4} and X-ray \citep{karpova2023} fluxes of \src, 
$F_\gamma = 4.2\times 10^{-12}$~\flux\ and $F_X = 1.7 \times 10^{-13}$~\flux, 
we estimated the corresponding luminosities to be $L_\gamma = 6.5 \times 10^{33}$~\ergs\ 
and $L_X = 2.6 \times 10^{32}$~\ergs\ at the distance of 3.6~kpc.
These values are consistent with those observed for other RB systems \citep{spidercat}.

Our light curve modelling, together with the derived mass functions from the radio and optical spectroscopy, suggests a potentially high-mass pulsar with $M_{\rm p} = 2.12(11)$~\msun.
Such an object could shed light on the equation of state of superdense matter inside neutron stars \citep[e.g.,][]{Romani2026}. 
The companion mass is quite low, $\approx 0.3$~\msun. 
The latter is roughly a factor of two smaller than that expected for a zero-age main-sequence star of equivalent radius, confirming that the secondary is substantially evolved and has lost a significant fraction of its initial mass. 
This picture aligns perfectly with the fact that the pulsar has been spun up to an exceptionally short period, indicating that it has successfully accreted a substantial amount of matter and gained significant mass during the recycling phase.

New optical observations are necessary to further investigate the long-term variability 
of the \src\ light curve. 
If the cold spot disappears, the clean, full orbital-cycle light curves will provide much tighter geometric constraints, thereby confirming the high mass of the pulsar. 
Furthermore, a comprehensive radio and $\gamma$-ray timing analysis will be presented in a future work.

%%%%%%%%%%%%%%%%%%%%%%%%%%%%%%%%%%%%%%%%%%%%%%%%%%%%%%%%
%%%%%%%%%%%%%%%%%%%%%%%%%%%%%%%%%%%%%%%%%%%%%%%%%%%%%%%%
%%%%%%%%%%%%%%%%%%%%%%%%%%%%%%%%%%%%%%%%%%%%%%%%%%%%%%%%
%%%%%%%%%%%%%%%%%%%%%%%%%%%%%%%%%%%%%%%%%%%%%%%%%%%%%%%%

\section*{Acknowledgements}

We thank the anonymous referee for their useful and constructive comments.
The work is based on observations carried out at the Gran Telescopio Canarias (GTC), installed at the Spanish Observatorio del Roque de los Muchachos of the Instituto de Astrof\'isica de Canarias, in the island of La Palma, and
the Observatorio Astron\'omico Nacional on the Sierra San Pedro M\'artir (OAN-SPM), Baja California, M\'exico. 
We thank the daytime and night support staff at the OAN-SPM for facilitating and helping obtain our observations.
The National Radio Astronomy Observatory and Green Bank Observatory (GBO) are facilities of the U.S. National Science Foundation operated under cooperative agreement by Associated Universities, Inc.
We thank our GBO `project friend' Ryan Lynch as well as the GBT telescope operators for their help in planning and executing our observations.
Computational resources for the radio-domain search were provided by the WVU Research Computing Thorny Flat HPC cluster, partly funded by NSF OAC-1726534. 
This work has made use of data from the European Space Agency (ESA) mission {\it Gaia} (\url{https://www.cosmos.esa.int/gaia}), processed by the {\it Gaia} Data Processing and Analysis Consortium (DPAC, \url{https://www.cosmos.esa.int/web/gaia/dpac/consortium}). Funding for the DPAC has been provided by national institutions, in particular the institutions participating in the {\it Gaia} Multilateral Agreement. 
The Pan-STARRS1 Surveys (PS1) and the PS1 public science archive have been made possible through contributions by the Institute for Astronomy, the University of Hawaii, the Pan-STARRS Project Office, the Max-Planck Society and its participating institutes, the Max Planck Institute for Astronomy, Heidelberg and the Max Planck Institute for Extraterrestrial Physics, Garching, The Johns Hopkins University, Durham University, the University of Edinburgh, the Queen's University Belfast, the Harvard-Smithsonian Center for Astrophysics, the Las Cumbres Observatory Global Telescope Network Incorporated, the National Central University of Taiwan, the Space Telescope Science Institute, the National Aeronautics and Space Administration under Grant No. NNX08AR22G issued through the Planetary Science Division of the NASA Science Mission Directorate, the National Science Foundation Grant No. AST-1238877, the University of Maryland, Eotvos Lorand University (ELTE), the Los Alamos National Laboratory, and the Gordon and Betty Moore Foundation.
The work of A.V.K., A.V.B. and D.A.Z. (optical data reduction, RVs measurements) was supported by the baseline project FFUG-2024-0002 of the Ioffe Institute. A.K. acknowledges the DGAPA-PAPIIT grant IN119426. S.V.Z. acknowledges the DGAPA-PAPIIT grant IN105826. M.L.M acknowledges the DGAPA-PAPIIT grant IA105024.
J.D.T acknowledges support from the NASA West Virginia Space Grant Consortium, Grant~\#~80NSSC25M7079.
E.~F. is supported by the National Science Foundation under grant AST-2407399.
D.A.Z. thanks Pirinem School of Theoretical Physics for hospitality.
%J.D.T. dedicates the radio discovery of PSR J2053+6905 to the memory of Jade, his beloved feline companion. 

%%%%%%%%%%%%%%%%%%%%%%%%%%%%%%%%%%%%%%%%%%%%%%%%%%%%%%%%
%%%%%%%%%%%%%%%%%%%%%%%%%%%%%%%%%%%%%%%%%%%%%%%%%%%%%%%%
%%%%%%%%%%%%%%%%%%%%%%%%%%%%%%%%%%%%%%%%%%%%%%%%%%%%%%%%
%%%%%%%%%%%%%%%%%%%%%%%%%%%%%%%%%%%%%%%%%%%%%%%%%%%%%%%%

\section*{Data Availability}

The GTC data are available through the archive \url{https://gtc.sdc.cab.inta-csic.es/gtc/} and the OAN-SPM and MAO data -- on request. The raw GBT data will become publicly available on the NRAO data archive after 9 January 2027.

\bibliographystyle{mnras}
\bibliography{ref}

\begin{thebibliography}{}
\makeatletter
\relax
\def\mn@urlcharsother{\let\do\@makeother \do\$\do\&\do\#\do\^\do\_\do\%\do\~}
\def\mn@doi{\begingroup\mn@urlcharsother \@ifnextchar [ {\mn@doi@}
  {\mn@doi@[]}}
\def\mn@doi@[#1]#2{\def\@tempa{#1}\ifx\@tempa\@empty \href
  {http://dx.doi.org/#2} {doi:#2}\else \href {http://dx.doi.org/#2} {#1}\fi
  \endgroup}
\def\mn@eprint#1#2{\mn@eprint@#1:#2::\@nil}
\def\mn@eprint@arXiv#1{\href {http://arxiv.org/abs/#1} {{\tt arXiv:#1}}}
\def\mn@eprint@dblp#1{\href {http://dblp.uni-trier.de/rec/bibtex/#1.xml}
  {dblp:#1}}
\def\mn@eprint@#1:#2:#3:#4\@nil{\def\@tempa {#1}\def\@tempb {#2}\def\@tempc
  {#3}\ifx \@tempc \@empty \let \@tempc \@tempb \let \@tempb \@tempa \fi \ifx
  \@tempb \@empty \def\@tempb {arXiv}\fi \@ifundefined
  {mn@eprint@\@tempb}{\@tempb:\@tempc}{\expandafter \expandafter \csname
  mn@eprint@\@tempb\endcsname \expandafter{\@tempc}}}

\bibitem[\protect\citeauthoryear{{Andersen} \& {Ransom}}{{Andersen} \&
  {Ransom}}{2018}]{Andersen2018_jerk}
{Andersen} B.~C.,  {Ransom} S.~M.,  2018, \mn@doi [\apjl]
  {10.3847/2041-8213/aad59f}, \href
  {https://ui.adsabs.harvard.edu/abs/2018ApJ...863L..13A} {863, L13}

\bibitem[\protect\citeauthoryear{{Ballet}, {Bruel}, {Burnett}, {Lott}  \& {The
  Fermi-LAT collaboration}}{{Ballet} et~al.}{2023}]{4fgl-dr4}
{Ballet} J.,  {Bruel} P.,  {Burnett} T.~H.,  {Lott} B.,   {The Fermi-LAT
  collaboration} 2023, \mn@doi [arXiv e-prints] {10.48550/arXiv.2307.12546},
  \href {https://ui.adsabs.harvard.edu/abs/2023arXiv230712546B} {p.
  arXiv:2307.12546}

\bibitem[\protect\citeauthoryear{{Bangale} et~al.,}{{Bangale}
  et~al.}{2024}]{Bangale2024}
{Bangale} P.,  et~al., 2024, \mn@doi [\apj] {10.3847/1538-4357/ad2994}, \href
  {https://ui.adsabs.harvard.edu/abs/2024ApJ...966..161B} {966, 161}

\bibitem[\protect\citeauthoryear{{Braglia} et~al.,}{{Braglia}
  et~al.}{2020}]{braglia2020}
{Braglia} C.,  et~al., 2020, \mn@doi [\mnras] {10.1093/mnras/staa2339}, \href
  {https://ui.adsabs.harvard.edu/abs/2020MNRAS.497.5364B} {497, 5364}

\bibitem[\protect\citeauthoryear{{CHIME/Pulsar Collaboration}
  et~al.,}{{CHIME/Pulsar Collaboration} et~al.}{2021}]{CHIME2021}
{CHIME/Pulsar Collaboration} et~al., 2021, \mn@doi [\apjs]
  {10.3847/1538-4365/abfdcb}, \href
  {https://ui.adsabs.harvard.edu/abs/2021ApJS..255....5C} {255, 5}

\bibitem[\protect\citeauthoryear{{Clark} et~al.,}{{Clark}
  et~al.}{2021}]{clark2021}
{Clark} C.~J.,  et~al., 2021, \mn@doi [\mnras] {10.1093/mnras/staa3484}, \href
  {https://ui.adsabs.harvard.edu/abs/2021MNRAS.502..915C} {502, 915}

\bibitem[\protect\citeauthoryear{{Clark} et~al.,}{{Clark}
  et~al.}{2023}]{Clark2023}
{Clark} C.~J.,  et~al., 2023, \mn@doi [\mnras] {10.1093/mnras/stac3742}, \href
  {https://ui.adsabs.harvard.edu/abs/2023MNRAS.519.5590C} {519, 5590}

\bibitem[\protect\citeauthoryear{{Cordes} \& {Lazio}}{{Cordes} \&
  {Lazio}}{2002}]{ne2001}
{Cordes} J.~M.,  {Lazio} T.~J.~W.,  2002, preprint, \href
  {https://ui.adsabs.harvard.edu/abs/2002astro.ph..7156C} {} (\mn@eprint {}
  {astro-ph/0207156})

\bibitem[\protect\citeauthoryear{{Flewelling} et~al.,}{{Flewelling}
  et~al.}{2020}]{flewelling2020}
{Flewelling} H.~A.,  et~al., 2020, \mn@doi [\apjs] {10.3847/1538-4365/abb82d},
  \href {https://ui.adsabs.harvard.edu/abs/2020ApJS..251....7F} {251, 7}

\bibitem[\protect\citeauthoryear{{Gaia Collaboration} et~al.,}{{Gaia
  Collaboration} et~al.}{2016}]{gaia2016}
{Gaia Collaboration} et~al., 2016, \mn@doi [\aap]
  {10.1051/0004-6361/201629272}, \href
  {https://ui.adsabs.harvard.edu/abs/2016A&A...595A...1G} {595, A1}

\bibitem[\protect\citeauthoryear{{Gaia Collaboration} et~al.,}{{Gaia
  Collaboration} et~al.}{2023}]{gaia-dr3}
{Gaia Collaboration} et~al., 2023, \mn@doi [\aap]
  {10.1051/0004-6361/202243940}, \href
  {https://ui.adsabs.harvard.edu/abs/2023A&A...674A...1G} {674, A1}

\bibitem[\protect\citeauthoryear{{Hessels}, {Ransom}, {Stairs}, {Freire},
  {Kaspi}  \& {Camilo}}{{Hessels} et~al.}{2006}]{Hessels2006}
{Hessels} J. W.~T.,  {Ransom} S.~M.,  {Stairs} I.~H.,  {Freire} P. C.~C.,
  {Kaspi} V.~M.,   {Camilo} F.,  2006, \mn@doi [Science]
  {10.1126/science.1123430}, \href
  {https://ui.adsabs.harvard.edu/abs/2006Sci...311.1901H} {311, 1901}

\bibitem[\protect\citeauthoryear{{Hobbs}, {Lorimer}, {Lyne}  \&
  {Kramer}}{{Hobbs} et~al.}{2005}]{hobbs2005}
{Hobbs} G.,  {Lorimer} D.~R.,  {Lyne} A.~G.,   {Kramer} M.,  2005, \mn@doi
  [\mnras] {10.1111/j.1365-2966.2005.09087.x}, \href
  {https://ui.adsabs.harvard.edu/abs/2005MNRAS.360..974H} {360, 974}

\bibitem[\protect\citeauthoryear{{Kandel} \& {Romani}}{{Kandel} \&
  {Romani}}{2020}]{kandel&romani2020}
{Kandel} D.,  {Romani} R.~W.,  2020, \mn@doi [\apj] {10.3847/1538-4357/ab7b62},
  \href {https://ui.adsabs.harvard.edu/abs/2020ApJ...892..101K} {892, 101}

\bibitem[\protect\citeauthoryear{{Kandel}, {Romani}, {Filippenko}, {Brink}  \&
  {Zheng}}{{Kandel} et~al.}{2020}]{kandel2020}
{Kandel} D.,  {Romani} R.~W.,  {Filippenko} A.~V.,  {Brink} T.~G.,   {Zheng}
  W.,  2020, \mn@doi [\apj] {10.3847/1538-4357/abb6fd}, \href
  {https://ui.adsabs.harvard.edu/abs/2020ApJ...903...39K} {903, 39}

\bibitem[\protect\citeauthoryear{{Karpova}, {Zyuzin}, {Shibanov}  \&
  {Gilfanov}}{{Karpova} et~al.}{2023}]{karpova2023}
{Karpova} A.~V.,  {Zyuzin} D.~A.,  {Shibanov} Y.~A.,   {Gilfanov} M.~R.,  2023,
  \mn@doi [\mnras] {10.1093/mnras/stad1992}, \href
  {https://ui.adsabs.harvard.edu/abs/2023MNRAS.524.3020K} {524, 3020}

\bibitem[\protect\citeauthoryear{{Kerby} et~al.,}{{Kerby}
  et~al.}{2021}]{kerby2021}
{Kerby} S.,  et~al., 2021, \mn@doi [\apj] {10.3847/1538-4357/ac2e91}, \href
  {https://ui.adsabs.harvard.edu/abs/2021ApJ...923...75K} {923, 75}

\bibitem[\protect\citeauthoryear{{Kesseli}, {West}, {Veyette}, {Harrison},
  {Feldman}  \& {Bochanski}}{{Kesseli} et~al.}{2017}]{Kesseli2017}
{Kesseli} A.~Y.,  {West} A.~A.,  {Veyette} M.,  {Harrison} B.,  {Feldman} D.,
  {Bochanski} J.~J.,  2017, \mn@doi [\apjs] {10.3847/1538-4365/aa656d}, \href
  {https://ui.adsabs.harvard.edu/abs/2017ApJS..230...16K} {230, 16}

\bibitem[\protect\citeauthoryear{{Kiziltan}, {Kottas}, {De Yoreo}  \&
  {Thorsett}}{{Kiziltan} et~al.}{2013}]{Kiziltan2013}
{Kiziltan} B.,  {Kottas} A.,  {De Yoreo} M.,   {Thorsett} S.~E.,  2013, \mn@doi
  [\apj] {10.1088/0004-637X/778/1/66}, \href
  {https://ui.adsabs.harvard.edu/abs/2013ApJ...778...66K} {778, 66}

\bibitem[\protect\citeauthoryear{{Koljonen} \& {Linares}}{{Koljonen} \&
  {Linares}}{2025}]{spidercat}
{Koljonen} K. I.~I.,  {Linares} M.,  2025, \mn@doi [\apj]
  {10.3847/1538-4357/ae08a5}, \href
  {https://ui.adsabs.harvard.edu/abs/2025ApJ...994....8K} {994, 8}

\bibitem[\protect\citeauthoryear{{Kostov} \& {Bonev}}{{Kostov} \&
  {Bonev}}{2018}]{2018BlgAJ..28....3K}
{Kostov} A.,  {Bonev} T.,  2018, \mn@doi [Bulgarian Astronomical Journal]
  {10.48550/arXiv.1706.06147}, \href
  {https://ui.adsabs.harvard.edu/abs/2018BlgAJ..28....3K} {28, 3}

\bibitem[\protect\citeauthoryear{{Kumar}, {Kumar}, {Thakur}, {Kumar}, {Kumar},
  {Sharma}, {Agrawal}  \& {Dhiman}}{{Kumar} et~al.}{2023}]{kumar2023}
{Kumar} R.,  {Kumar} M.,  {Thakur} V.,  {Kumar} S.,  {Kumar} P.,  {Sharma} A.,
  {Agrawal} B.~K.,   {Dhiman} S.~K.,  2023, \mn@doi [\prc]
  {10.1103/PhysRevC.107.055805}, \href
  {https://ui.adsabs.harvard.edu/abs/2023PhRvC.107e5805K} {107, 055805}

\bibitem[\protect\citeauthoryear{{Lampton}, {Margon}  \& {Bowyer}}{{Lampton}
  et~al.}{1976}]{1976ApJ...208..177L}
{Lampton} M.,  {Margon} B.,   {Bowyer} S.,  1976, \mn@doi [\apj]
  {10.1086/154592}, \href
  {https://ui.adsabs.harvard.edu/abs/1976ApJ...208..177L} {208, 177}

\bibitem[\protect\citeauthoryear{{Linares}, {Shahbaz}  \& {Casares}}{{Linares}
  et~al.}{2018}]{Linares2018}
{Linares} M.,  {Shahbaz} T.,   {Casares} J.,  2018, \mn@doi [\apj]
  {10.3847/1538-4357/aabde6}, \href
  {https://ui.adsabs.harvard.edu/abs/2018ApJ...859...54L} {859, 54}

\bibitem[\protect\citeauthoryear{{Lorimer} \& {Kramer}}{{Lorimer} \&
  {Kramer}}{2004}]{handbook2004}
{Lorimer} D.~R.,  {Kramer} M.,  2004, {Handbook of Pulsar Astronomy}.
Cambridge University Press

\bibitem[\protect\citeauthoryear{{Manchester}, {Hobbs}, {Teoh}  \&
  {Hobbs}}{{Manchester} et~al.}{2005}]{atnf}
{Manchester} R.~N.,  {Hobbs} G.~B.,  {Teoh} A.,   {Hobbs} M.,  2005, \mn@doi
  [\aj] {10.1086/428488}, \href
  {https://ui.adsabs.harvard.edu/abs/2005AJ....129.1993M} {129, 1993}

\bibitem[\protect\citeauthoryear{{Mata S{\'a}nchez} et~al.,}{{Mata S{\'a}nchez}
  et~al.}{2023}]{matasanchez2023}
{Mata S{\'a}nchez} D.,  et~al., 2023, \mn@doi [\mnras] {10.1093/mnras/stad203},
  \href {https://ui.adsabs.harvard.edu/abs/2023MNRAS.tmp..227M} {520, 2217}

\bibitem[\protect\citeauthoryear{{Moiseev}}{{Moiseev}}{2008}]{Moiseev2008}
{Moiseev} A.~V.,  2008, \mn@doi [Astrophysical Bulletin]
  {10.1007/s11755-008-1007-5}, \href
  {https://ui.adsabs.harvard.edu/abs/2008AstBu..63...70M} {63, 70}

\bibitem[\protect\citeauthoryear{{Ocker} \& {Cordes}}{{Ocker} \&
  {Cordes}}{2026}]{ne2025}
{Ocker} S.~K.,  {Cordes} J.~M.,  2026, \mn@doi [\apj]
  {10.3847/1538-4357/ae5825}, \href
  {https://ui.adsabs.harvard.edu/abs/2026ApJ..1002....3O} {1002, 3}

\bibitem[\protect\citeauthoryear{{Oke}}{{Oke}}{1990}]{Oke1990}
{Oke} J.~B.,  1990, \mn@doi [\aj] {10.1086/115444}, \href
  {https://ui.adsabs.harvard.edu/abs/1990AJ.....99.1621O} {99, 1621}

\bibitem[\protect\citeauthoryear{{Prochaska} et~al.,}{{Prochaska}
  et~al.}{2020a}]{pypeit:zenodo}
{Prochaska} J.~X.,  et~al., 2020a, {pypeit/PypeIt: Release 1.0.0},
  \mn@doi{10.5281/zenodo.3743493}

\bibitem[\protect\citeauthoryear{Prochaska et~al.,}{Prochaska
  et~al.}{2020b}]{pypeit:joss_pub}
Prochaska J.~X.,  et~al., 2020b, \mn@doi [Journal of Open Source Software]
  {10.21105/joss.02308}, 5, 2308

\bibitem[\protect\citeauthoryear{{Ransom}}{{Ransom}}{2011}]{presto2011}
{Ransom} S.,  2011, {PRESTO: PulsaR Exploration and Search TOolkit},
  Astrophysics Source Code Library, record ascl:1107.017 (\mn@eprint {ascl}
  {1107.017})

\bibitem[\protect\citeauthoryear{{Roberts}}{{Roberts}}{2013}]{roberts2013}
{Roberts} M. S.~E.,  2013, in {van Leeuwen} J.,  ed.,  IAU Symposium Vol. 291,
  Neutron Stars and Pulsars: Challenges and Opportunities after 80 years. pp
  127--132 (\mn@eprint {arXiv} {1210.6903}), \mn@doi{10.1017/S174392131202337X}

\bibitem[\protect\citeauthoryear{{Romani} \& {Sanchez}}{{Romani} \&
  {Sanchez}}{2016}]{romani&sanchez2016}
{Romani} R.~W.,  {Sanchez} N.,  2016, \mn@doi [\apj]
  {10.3847/0004-637X/828/1/7}, \href
  {https://ui.adsabs.harvard.edu/abs/2016ApJ...828....7R} {828, 7}

\bibitem[\protect\citeauthoryear{{Romani}, {Beleznay}, {Filippenko}, {Brink}
  \& {Zheng}}{{Romani} et~al.}{2026}]{Romani2026}
{Romani} R.~W.,  {Beleznay} M.,  {Filippenko} A.~V.,  {Brink} T.~G.,   {Zheng}
  W.,  2026, \mn@doi [\apj] {10.3847/1538-4357/ae28c5}, \href
  {https://ui.adsabs.harvard.edu/abs/2026ApJ...996..101R} {996, 101}

\bibitem[\protect\citeauthoryear{{Rosenthal} et~al.,}{{Rosenthal}
  et~al.}{2025}]{Rosenthal2025eclipses}
{Rosenthal} A.~C.,  et~al., 2025, \mn@doi [\apj] {10.3847/1538-4357/adb8cd},
  \href {https://ui.adsabs.harvard.edu/abs/2025ApJ...982..170R} {982, 170}

\bibitem[\protect\citeauthoryear{{Sanchez} \& {Romani}}{{Sanchez} \&
  {Romani}}{2017}]{sanchez&romani2017}
{Sanchez} N.,  {Romani} R.~W.,  2017, \mn@doi [\apj]
  {10.3847/1538-4357/aa7a02}, \href
  {https://ui.adsabs.harvard.edu/abs/2017ApJ...845...42S} {845, 42}

\bibitem[\protect\citeauthoryear{{Schuster} \& {Parrao}}{{Schuster} \&
  {Parrao}}{2001}]{schuster2}
{Schuster} W.~J.,  {Parrao} L.,  2001, \rmxaa, \href
  {https://ui.adsabs.harvard.edu/abs/2001RMxAA..37..187S} {37, 187}

\bibitem[\protect\citeauthoryear{{Schuster}, {Parrao}  \&
  {Guichard}}{{Schuster} et~al.}{2002}]{schuster1}
{Schuster} W.~J.,  {Parrao} L.,   {Guichard} J.,  2002, Journal of Astronomical
  Data, \href {https://ui.adsabs.harvard.edu/abs/2002JAD.....8....2S} {8, 2}

\bibitem[\protect\citeauthoryear{{Shamohammadi} et~al.,}{{Shamohammadi}
  et~al.}{2024}]{Shamohammadi2024pulsarvel}
{Shamohammadi} M.,  et~al., 2024, \mn@doi [\mnras] {10.1093/mnras/stae016},
  \href {https://ui.adsabs.harvard.edu/abs/2024MNRAS.530..287S} {530, 287}

\bibitem[\protect\citeauthoryear{{Strader} et~al.,}{{Strader}
  et~al.}{2019}]{strader2019}
{Strader} J.,  et~al., 2019, \mn@doi [\apj] {10.3847/1538-4357/aafbaa}, \href
  {https://ui.adsabs.harvard.edu/abs/2019ApJ...872...42S} {872, 42}

\bibitem[\protect\citeauthoryear{{Taylor}, {Ransom}  \& {Padmanabh}}{{Taylor}
  et~al.}{2024}]{Taylor2024APTB}
{Taylor} J.,  {Ransom} S.,   {Padmanabh} P.~V.,  2024, \mn@doi [\apj]
  {10.3847/1538-4357/ad1ce9}, \href
  {https://ui.adsabs.harvard.edu/abs/2024ApJ...964..128T} {964, 128}

\bibitem[\protect\citeauthoryear{{Thongmeearkom} et~al.,}{{Thongmeearkom}
  et~al.}{2026}]{Thongmeearkom2026}
{Thongmeearkom} T.,  et~al., 2026, \mn@doi [\mnras] {10.1093/mnras/stag406},
  \href {https://ui.adsabs.harvard.edu/abs/2026MNRAS.547ag406T} {547, stag406}

\bibitem[\protect\citeauthoryear{{Turchetta}, {Linares}, {Koljonen}  \&
  {Sen}}{{Turchetta} et~al.}{2023}]{Turchetta2023}
{Turchetta} M.,  {Linares} M.,  {Koljonen} K.,   {Sen} B.,  2023, \mn@doi
  [\mnras] {10.1093/mnras/stad2435}, \href
  {https://ui.adsabs.harvard.edu/abs/2023MNRAS.525.2565T} {525, 2565}

\bibitem[\protect\citeauthoryear{{Turner} et~al.,}{{Turner}
  et~al.}{2024}]{Turner2024scint}
{Turner} J.~E.,  et~al., 2024, \mn@doi [\apj] {10.3847/1538-4357/ad5af9}, \href
  {https://ui.adsabs.harvard.edu/abs/2024ApJ...972...16T} {972, 16}

\bibitem[\protect\citeauthoryear{{Voisin}, {Kennedy}, {Breton}, {Clark}  \&
  {Mata-S{\'a}nchez}}{{Voisin} et~al.}{2020}]{voisin2020}
{Voisin} G.,  {Kennedy} M.~R.,  {Breton} R.~P.,  {Clark} C.~J.,
  {Mata-S{\'a}nchez} D.,  2020, \mn@doi [\mnras] {10.1093/mnras/staa2876},
  \href {https://ui.adsabs.harvard.edu/abs/2020MNRAS.499.1758V} {499, 1758}

\bibitem[\protect\citeauthoryear{{Wang} et~al.,}{{Wang}
  et~al.}{2025}]{wang2025}
{Wang} P.~F.,  et~al., 2025, \mn@doi [Research in Astronomy and Astrophysics]
  {10.1088/1674-4527/ada3b8}, \href
  {https://ui.adsabs.harvard.edu/abs/2025RAA....25a4003W} {25, 014003}

\bibitem[\protect\citeauthoryear{{Yao}, {Manchester}  \& {Wang}}{{Yao}
  et~al.}{2017}]{ymw2016}
{Yao} J.~M.,  {Manchester} R.~N.,   {Wang} N.,  2017, \mn@doi [\apj]
  {10.3847/1538-4357/835/1/29}, \href
  {https://ui.adsabs.harvard.edu/abs/2017ApJ...835...29Y} {835, 29}

\bibitem[\protect\citeauthoryear{{Zharikov}, {Tovmassian}, {Aviles}, {Michel},
  {Gonzalez-Buitrago}  \& {Garc{\'\i}a-D{\'\i}az}}{{Zharikov}
  et~al.}{2013}]{zharikov2013}
{Zharikov} S.,  {Tovmassian} G.,  {Aviles} A.,  {Michel} R.,
  {Gonzalez-Buitrago} D.,   {Garc{\'\i}a-D{\'\i}az} M.~T.,  2013, \mn@doi
  [\aap] {10.1051/0004-6361/201220099}, \href
  {https://ui.adsabs.harvard.edu/abs/2013A&A...549A..77Z} {549, A77}

\bibitem[\protect\citeauthoryear{{Zharikov}, {Kirichenko}, {Zyuzin}, {Shibanov}
   \& {Deneva}}{{Zharikov} et~al.}{2019}]{zharikov2019}
{Zharikov} S.,  {Kirichenko} A.,  {Zyuzin} D.,  {Shibanov} Y.,   {Deneva}
  J.~S.,  2019, \mn@doi [\mnras] {10.1093/mnras/stz2475}, \href
  {https://ui.adsabs.harvard.edu/abs/2019MNRAS.489.5547Z} {489, 5547}

\bibitem[\protect\citeauthoryear{{de Martino} et~al.,}{{de Martino}
  et~al.}{2024}]{deMartino2024}
{de Martino} D.,  et~al., 2024, \mn@doi [\aap] {10.1051/0004-6361/202450955},
  \href {https://ui.adsabs.harvard.edu/abs/2024A&A...691A..36D} {691, A36}

\makeatother
\end{thebibliography}

% Don't change these lines
\bsp	% typesetting comment
\label{lastpage}
\end{document}